\documentclass[traditabstract,letter]{aa} 

\usepackage{graphicx}
\usepackage{natbib}
\usepackage{url}
\usepackage{multirow}
\usepackage{rotating}
\usepackage{txfonts}
\begin{document}

\newcommand{\zabs}{\ensuremath{z_{\rm abs}}}
\newcommand{\zem}{\ensuremath{z_{\rm em}}}
\newcommand{\dla}{damped Lyman-$\alpha$}
\newcommand{\Dla}{damped Lyman-$\alpha$}
\newcommand{\lya}{Ly-$\alpha$}
\newcommand{\lyb}{Ly-$\beta$}
\newcommand{\lyg}{Ly-$\gamma$}

\newcommand{\ArI}{\ion{Ar}{i}}
\newcommand{\CaII}{\ion{Ca}{ii}}
\newcommand{\CI}{\ion{C}{i}}
\newcommand{\CII}{\ion{C}{ii}}
\newcommand{\CIV}{\ion{C}{iv}}
\newcommand{\ClI}{\ion{Cl}{i}}
\newcommand{\ClII}{\ion{Cl}{ii}}
\newcommand{\CrII}{\ion{Cr}{ii}}
\newcommand{\CuII}{\ion{Cu}{ii}}
\newcommand{\DI}{\ion{D}{i}}
\newcommand{\FeI}{\ion{Fe}{i}}
\newcommand{\FeII}{\ion{Fe}{ii}}
\newcommand{\HI}{\ion{H}{i}}
\newcommand{\MgI}{\ion{Mg}{i}}
\newcommand{\MgII}{\ion{Mg}{ii}}
\newcommand{\MnII}{\ion{Mn}{ii}}
\newcommand{\NI}{\ion{N}{i}}
\newcommand{\NaI}{\ion{Na}{i}}
\newcommand{\NII}{\ion{N}{ii}}
\newcommand{\NV}{\ion{N}{v}}
\newcommand{\NiII}{\ion{Ni}{ii}}
\newcommand{\OI}{\ion{O}{i}}
\newcommand{\OII}{\ion{O}{ii}}
\newcommand{\OIII}{\ion{O}{iii}}
\newcommand{\OVI}{\ion{O}{vi}}
\newcommand{\PII}{\ion{P}{ii}}
\newcommand{\PbII}{\ion{Pb}{ii}}
\newcommand{\SI}{\ion{S}{i}}
\newcommand{\SII}{\ion{S}{ii}}
\newcommand{\SiII}{\ion{Si}{ii}}
\newcommand{\SiIV}{\ion{Si}{iv}}
\newcommand{\TiII}{\ion{Ti}{ii}}
\newcommand{\ZnII}{\ion{Zn}{ii}}
\newcommand{\AlII}{\ion{Al}{ii}}
\newcommand{\AlIII}{\ion{Al}{iii}}
\newcommand{\Ho}{\mbox{H$_0$}}
\newcommand{\angstrom}{\mbox{{\rm \AA}}}
\newcommand{\abs}[1]{\left| #1 \right|} 
\newcommand{\avg}[1]{\left< #1 \right>} 
\newcommand{\kms}{\ensuremath{{\rm km\,s^{-1}}}}
\newcommand{\cmsq}{\ensuremath{{\rm cm}^{-2}}}
\newcommand{\qso}{J1135$-$0010}
\newcommand{\qsolong}{SDSS\,J113520.39$-$001053.56}
\newcommand{\nhi}{n_{\rm HI}}

\newcommand{\iap}{CNRS-UPMC, UMR7095, Institut d'Astrophysique de Paris, 98bis bd Arago, 75014 Paris, France}
\newcommand{\uchile}{Departamento de Astronom\'ia, Universidad de Chile, 
Casilla 36-D, Santiago, Chile}
\newcommand{\eso}{European Southern Observatory, Alonso de C\'ordova 3107, 
Vitacura, Casilla 19001, Santiago 19, Chile}
\newcommand{\oat}{Osservatorio Astronomico di Trieste, Via G. B. Tiepolo 11, 34131 Trieste, Italy}

\title{Deuterium at high-redshift}
\subtitle{Primordial abundance in the $\zabs=2.621$ damped Ly-$\alpha$ system towards CTQ\,247
\thanks{Based on archival ESO data Prgm.~ID. 70.A-0017(A).}}
\titlerunning{Deuterium at high-redshift: Primordial abundance in the $\zabs = 2.621$ DLA towards CTQ\,247}
\author{P. Noterdaeme\inst{1}  \and S. L\'opez\inst{2} \and V. Dumont\inst{2} \and C. Ledoux\inst{3} \and 
P. Molaro\inst{4} \and P. Petitjean\inst{1}}

\institute{\iap\  -- \email{noterdaeme@iap.fr}
\and 
\uchile 
\and 
\eso 
\and 
\oat 
}
\date{}

   \abstract{The detection of neutral deuterium in the low-metallicity damped \mbox{Lyman-$\alpha$} system 
at $\zabs=2.621$ towards the quasar CTQ\,247 is reported. Using a high signal-to-noise and high spectral 
resolution ($R = 60\,000$) spectrum from the Very Large Telescope Ultraviolet and Visual Echelle 
Spectrograph, we precisely measure the deuterium-to-oxygen ratio $\log N(\DI)/N(\OI)=0.74\pm0.04$, as well 
as the overall oxygen abundance, $\log N(\OI)/N(\HI)=-5.29\pm0.10$ (or equivalently [O/H]~$=-1.99\pm0.10$ 
with respect to the solar value). Assuming uniform metallicity throughout the system, our measurement 
translates to (D/H)~$=(2.8^{+0.8}_{-0.6})\times10^{-5}$. This ratio is consistent within errors ($<0.4\,\sigma$) 
with the primordial ratio, (D/H)$_{\rm p} = (2.59\pm0.15)\times10^{-5}$, predicted by standard Big-Bang 
Nucleosynthesis using the WMAP7 value of the cosmological density of baryons 
($100\,\Omega_b h^2=2.249\pm0.056$). 
The \DI\ absorption lines are observed to be broader than the \OI\ absorption lines. From a consistent fit 
of the profiles we derive the turbulent broadening to be 5.2\,\kms\ and the temperature of the gas to be 
$T=8800\pm1500$~K, corresponding to a warm neutral medium.}


   \maketitle

\section{Introduction}

The primordial abundance of deuterium is fixed by Big-Bang Nucleosynthesis (BBN) and depends directly on the 
baryon-to-photon ratio $\eta$. The (D/H) ratio hence provides a baryometer of choice 
\citep[e.g.][]{Steigman07a}. Because of no (or only very marginal) subsequent production of deuterium 
\citep[e.g.][]{Epstein76,Prodanovic03}, and because this species is easily destroyed inside stars 
(astration), (D/H) is expected to decrease monotonically with cosmic time. To probe this 
scenario, much effort has been devoted to the measurement of the deuterium abundance, in particular in 
low-metallicity environments at high redshift ($z>2$). This is done from observations of \DI\ and \HI\ 
absorption lines in QSO absorption systems \citep{Adams76}. Unfortunately, measurements are not yet possible 
in the low density \lya\ forest because of the extremely low \DI\ column densities expected (there is a five 
order of magnitude difference in the abundance of the two hydrogen isotopes), so one has to turn to higher 
column density systems, namely Lyman-limit systems ($\log N(\HI)$~(cm$^{-2})>17$) and damped Lyman-$\alpha$ 
systems (DLAs, $\log N(\HI) \ge 20.3$). Although the latter are known to be related to star-forming regions, 
the deuterium abundance is expected to remain close to the primordial value in the low metallicity regime 
\citep[][]{Romano06}. The main problem is then the small velocity separation between \DI\ and \HI\ lines, 
$\Delta v \sim$~80~\kms, that causes \DI\ lines to be easily lost within much stronger \HI\ profiles. It may 
also be difficult to discern between \DI\ lines and unrelated absorptions from the \lya\ forest 
\citep[e.g.][]{Steigman94}. This explains why few robust measurements have been performed so far 
\citep[][]{Pettini08}.

In this letter, we present a new measurement of deuterium abundance in the DLA at $z_{\rm abs} = 2.621$ 
towards CTQ\,247 (also called Q\,0405$-$443 or J\,040718$-$441014). This system was first reported by 
\citet{Lopez01} from low resolution spectroscopy observations, together with two other DLAs along the same 
line of sight (at $\zabs=2.551$ and 2.595). \citet{Ledoux03} subsequently confirmed the DLAs from 
high-resolution data but focused mainly on the $\zabs = 2.595$ system, in which they detected absorption 
lines from molecular hydrogen. The metal abundances in the three DLAs were then studied by \citet{Lopez03}. 
We identified \DI\ absorption lines in the $z_{\rm abs} = 2.621$ system while screening the Very Large 
Telescope Ultraviolet and Visual Echelle Spectrograph (VLT/UVES) data archive. Interestingly, the system 
fulfils the optimal criteria for the determination of (D/H) as proposed by \citet{York02}: 
$N(\HI)$ above $10^{19}$\cmsq, warm gas (6\,000-10\,000~K), wide separation of the \DI-bearing component, 
and availability of \OI\ lines.

\section{Observations and data reduction \label{obs}}

CTQ\,247 has been observed several times in the 
past decade using VLT/UVES with different instrument setups and under
varying sky conditions. Here, we take advantage of UVES data
obtained through programme ID 70.A-0017(A) (P.I. Petitjean) using a
$0.8\arcsec$ slit width and 2$\times$2 CCD pixel binning. This homogeneous
dataset is composed of nine exposures totalling more than 12 hours of
integration time under good seeing conditions ($\sim\,0.8\arcsec$).
Dichroic \#1 was used to observe with the blue and red spectroscopic
arms simultaneously with central wavelengths adjusted to 390\,nm and
570\,nm respectively. Wavelength calibration frames were taken on
target using a ThAr lamp immediately after each science exposure. With
this setup, a resolving power of $R=60\,000$ was reached.
The spectra were reduced using the UVES data reduction 
pipeline v4.9.5 based on the ESO Common Pipeline Library v5.3.1.
Object and sky were optimally extracted 
using a properly over-sampled Gaussian or virtual profile depending 
on the S/N ratio\footnote{See ESO UVES pipeline user manual v19.}. 
This extraction is based on a generalisation of 
\citeauthor{Horne86}'s algorithm to
get combined optimal sky and object flux estimates via $\chi^2$-minimisation. 
Pixels affected by cosmic ray hits
and CCD defects were identified at the same time and removed via
$\kappa$-sigma clipping. Individual spectra were shifted to the
vacuum-heliocentric rest-frame before they were combined,
by weighting the flux in each pixel by the inverse of its variance.
The resulting spectrum has a S/N ratio per pixel of $\sim 55$ at $\lambda_{\rm
  obs}=4000$\,{\AA}.
Given the exquisite data quality, we were able to identify
a slight inter-order background over-subtraction, at the level of 1\%
at $\lambda_{\rm obs}<3800$\,{\AA} and increasing towards bluer wavelengths. We 
corrected this effet using the bottom of saturated \lya\ forest lines as reference for 
the true zero level.

\section{Analysis \label{analysis}}

\begin{figure}[!t]
\centering
\renewcommand{\tabcolsep}{0pt}
\begin{tabular}{cc}
\includegraphics[bb = 218 229 393 617,clip=,angle=90,width=0.46\hsize]{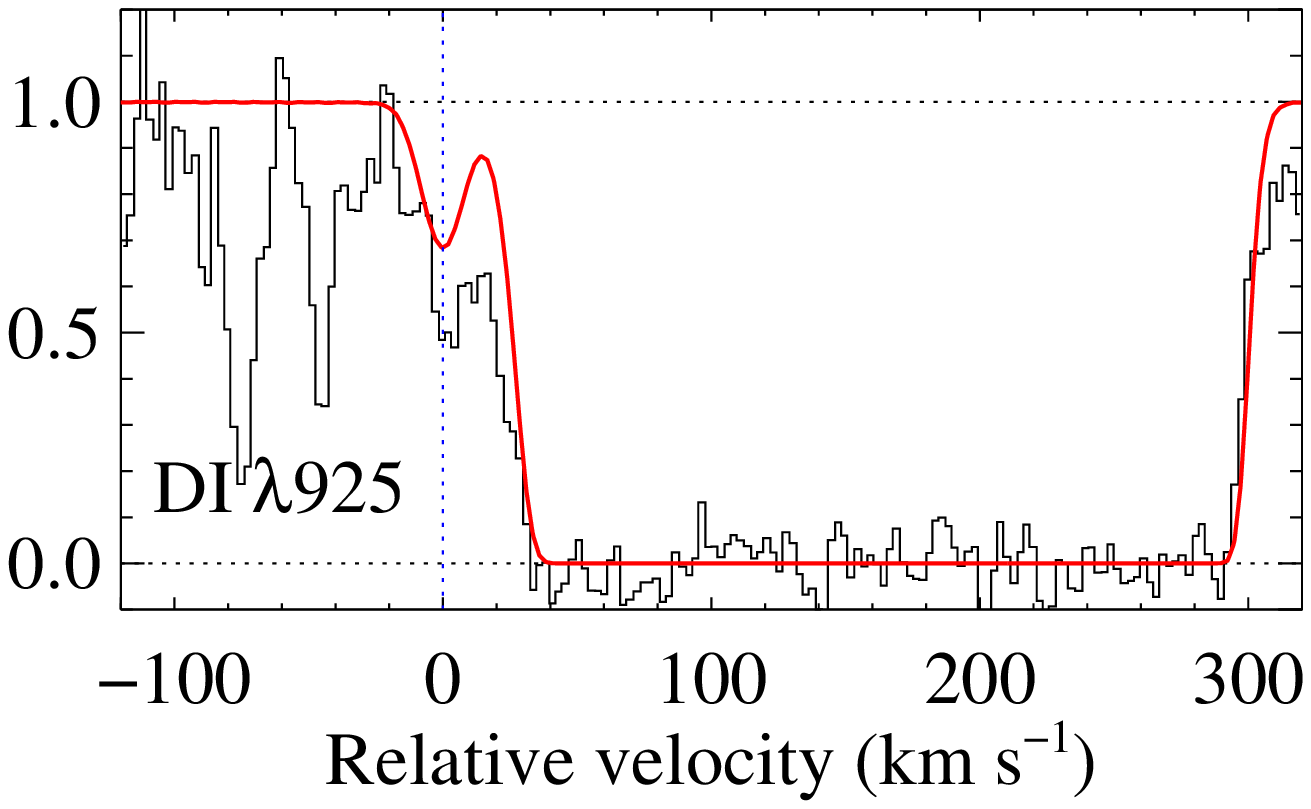}&
\includegraphics[bb = 218 229 393 617,clip=,angle=90,width=0.46\hsize]{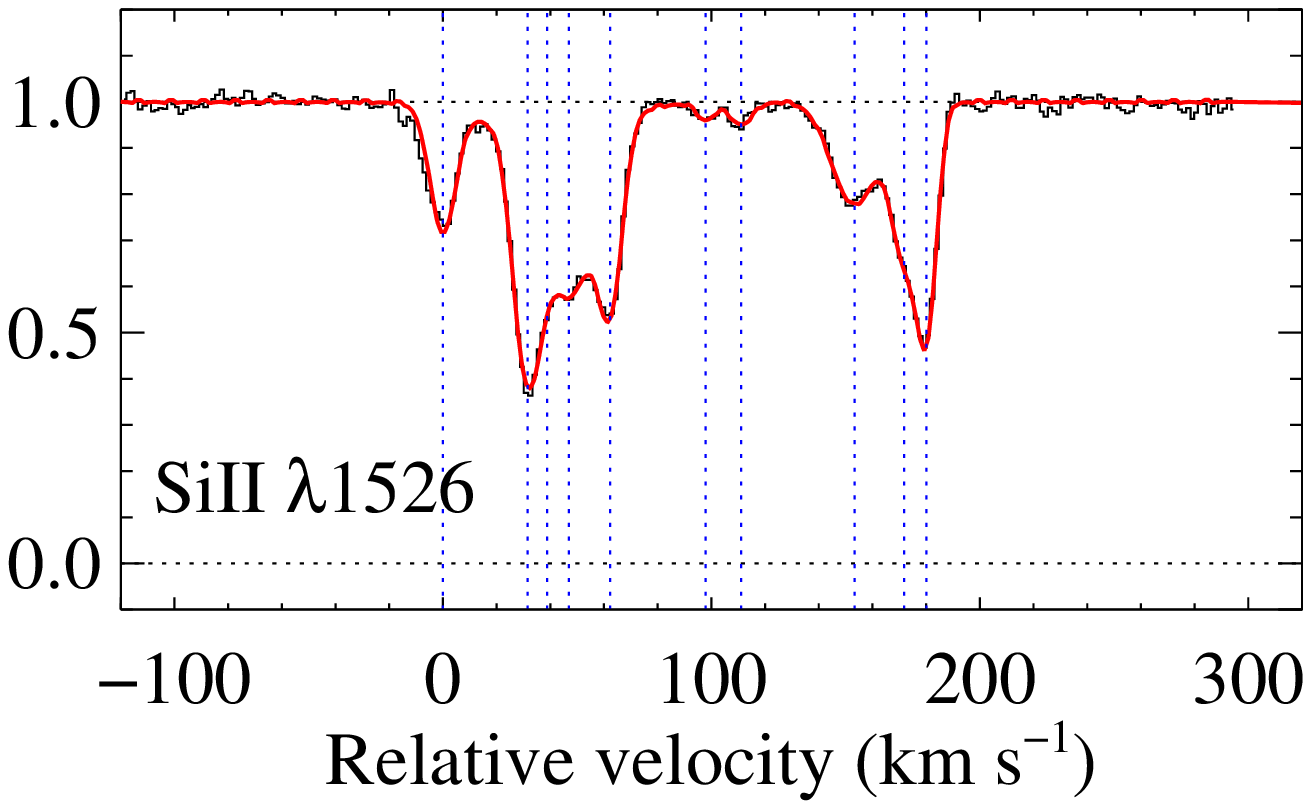}\\
\includegraphics[bb = 218 229 393 617,clip=,angle=90,width=0.46\hsize]{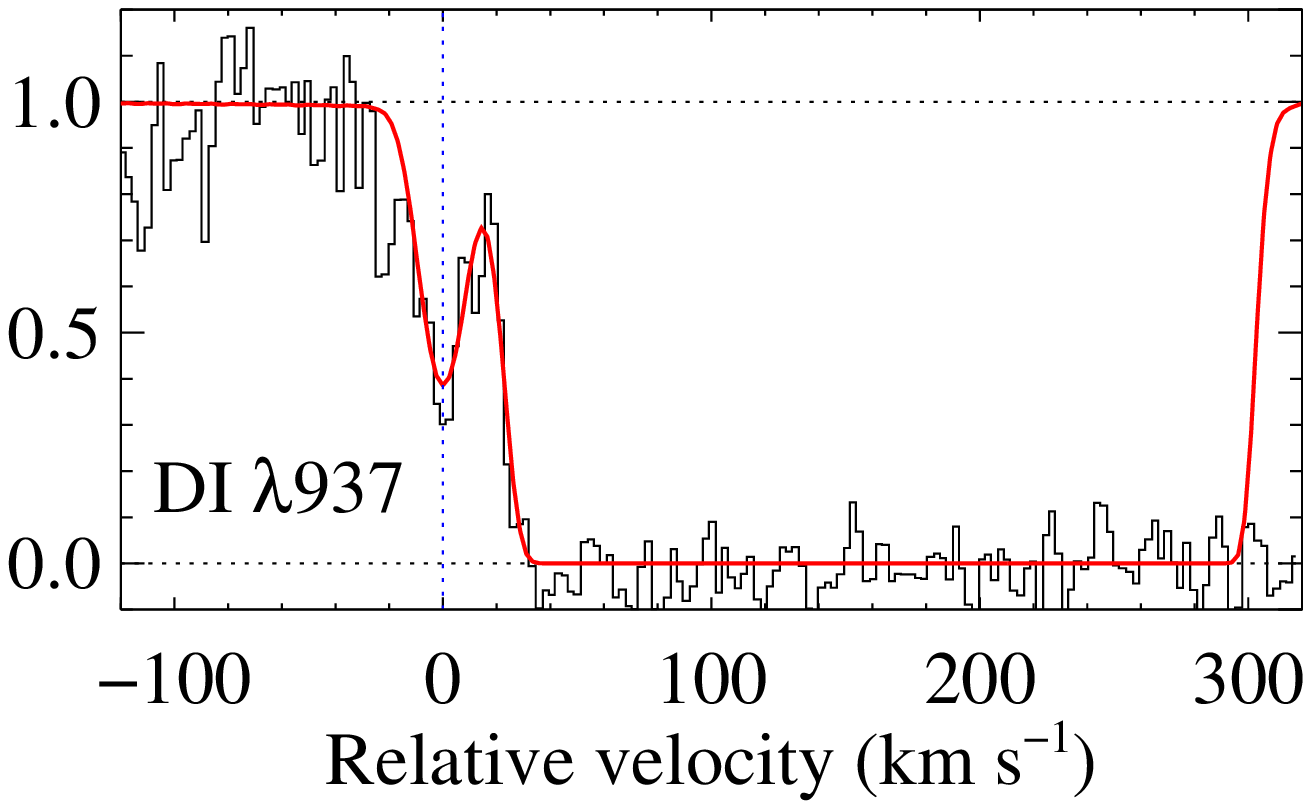}&
\includegraphics[bb = 218 229 393 617,clip=,angle=90,width=0.46\hsize]{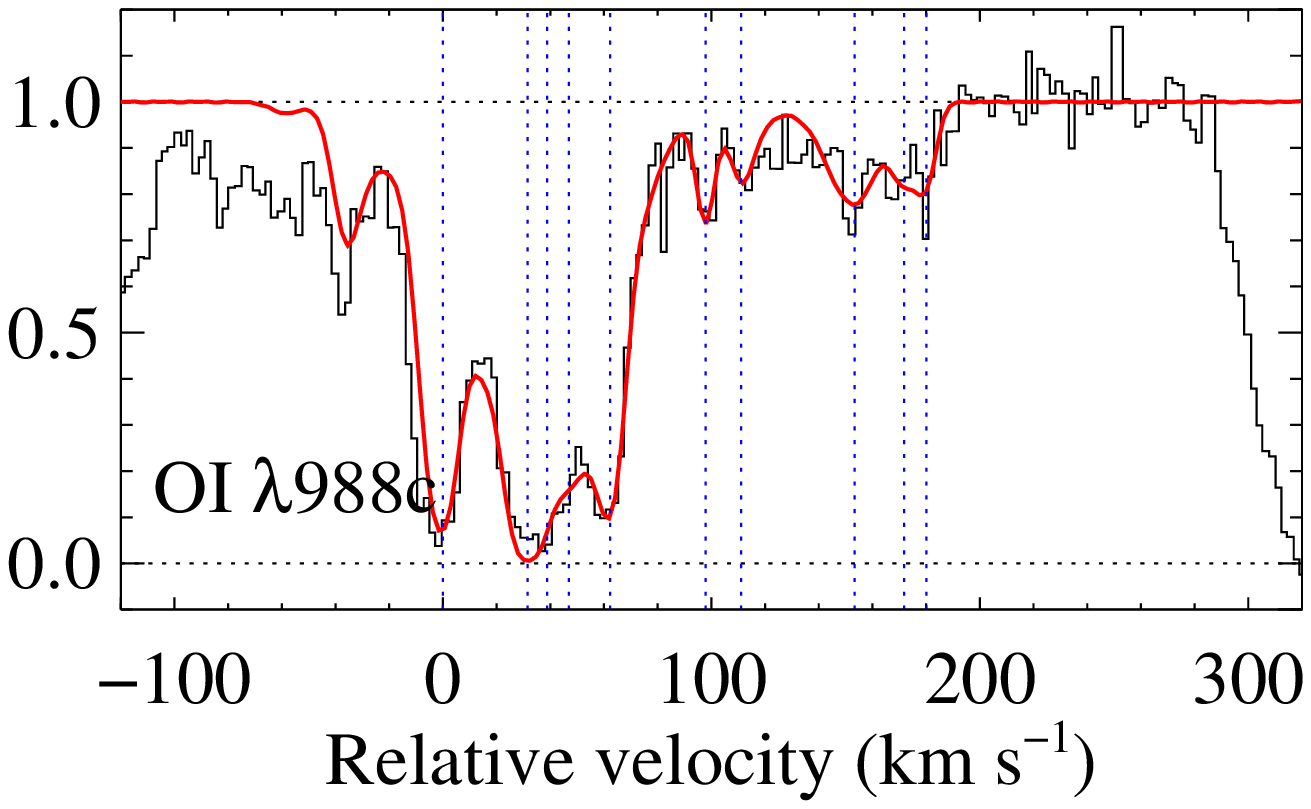}\\
\includegraphics[bb = 218 229 393 617,clip=,angle=90,width=0.46\hsize]{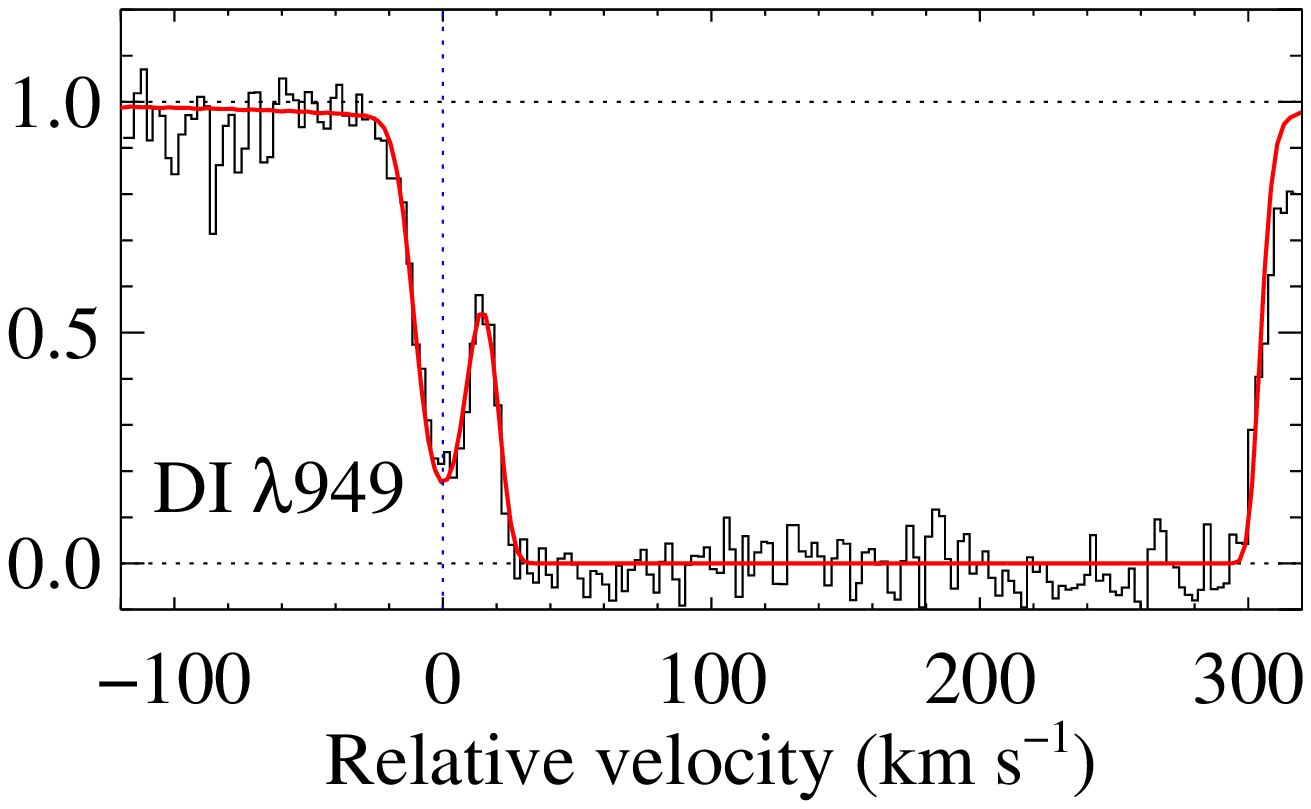}&
\includegraphics[bb = 218 229 393 617,clip=,angle=90,width=0.46\hsize]{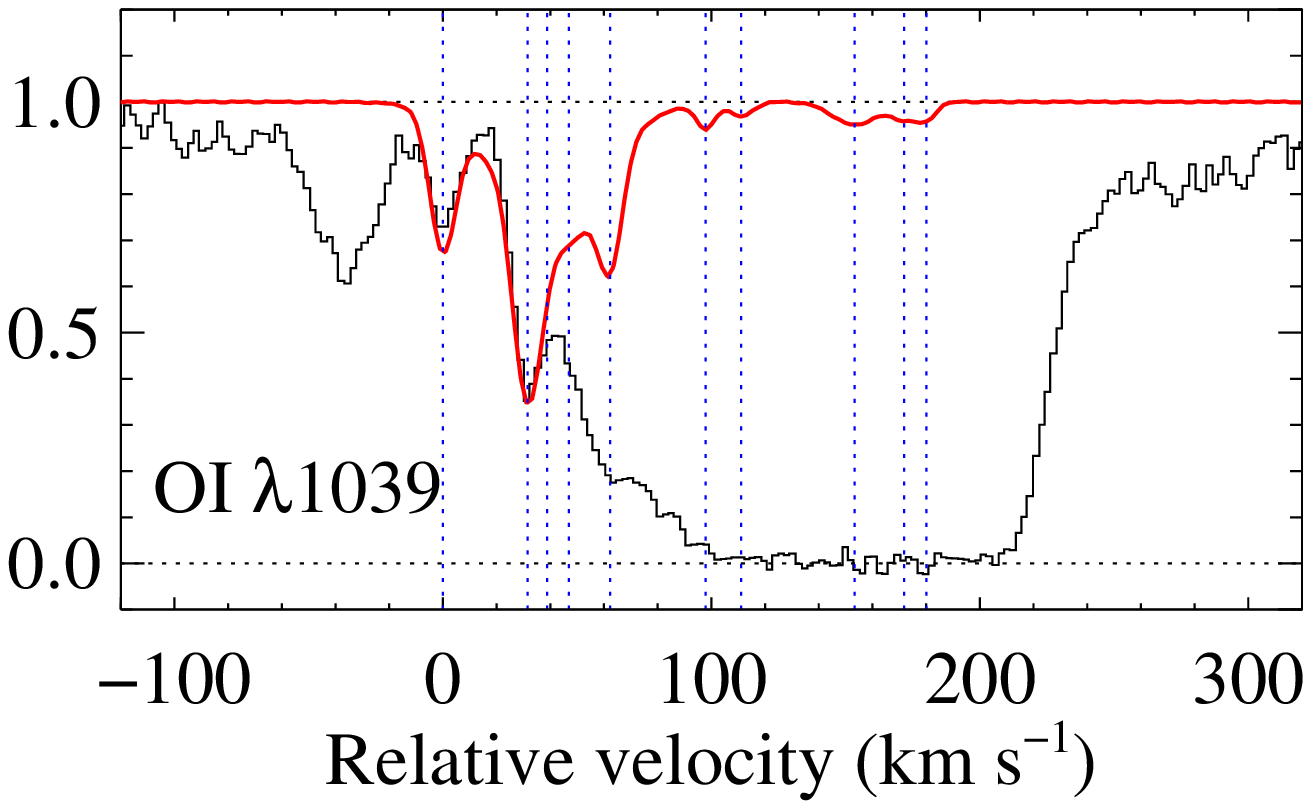}\\
\includegraphics[bb = 162 229 393 617,clip=,angle=90,width=0.46\hsize]{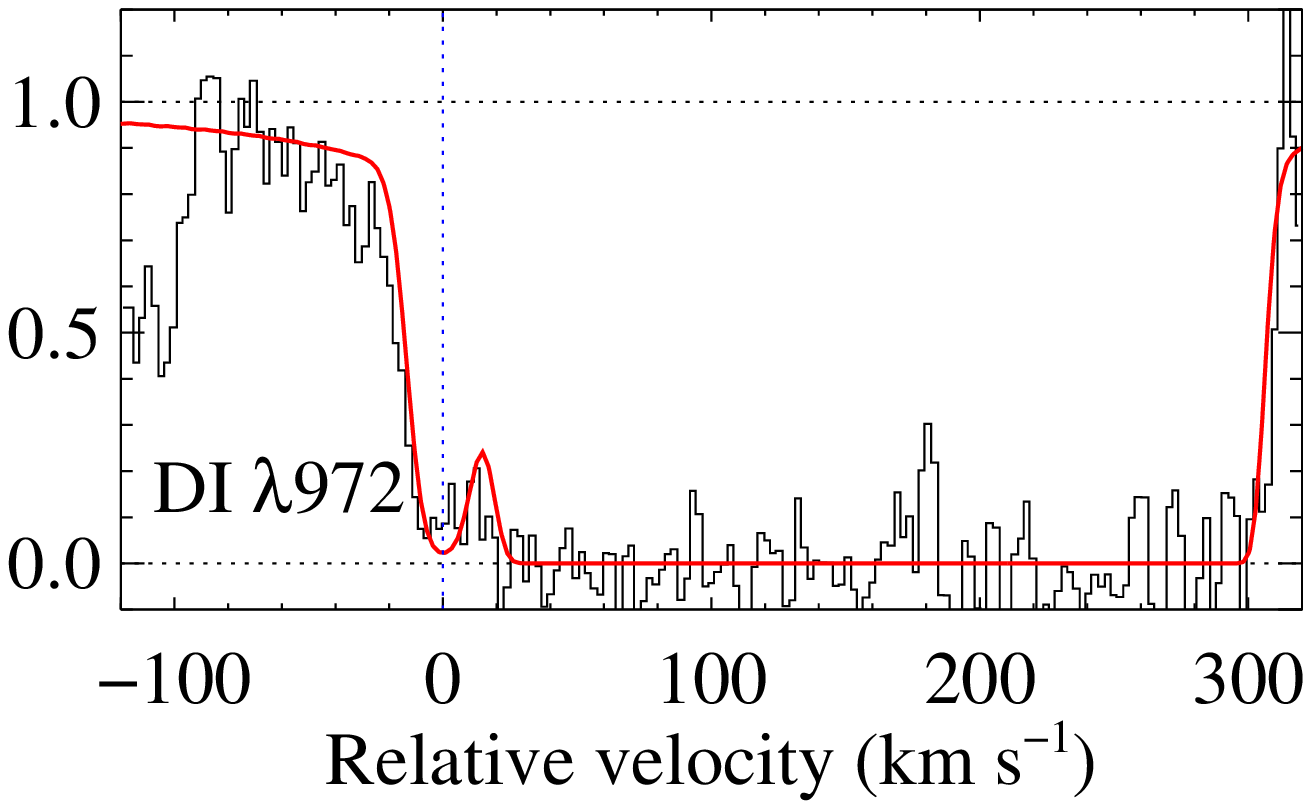}&
\includegraphics[bb = 162 229 393 617,clip=,angle=90,width=0.46\hsize]{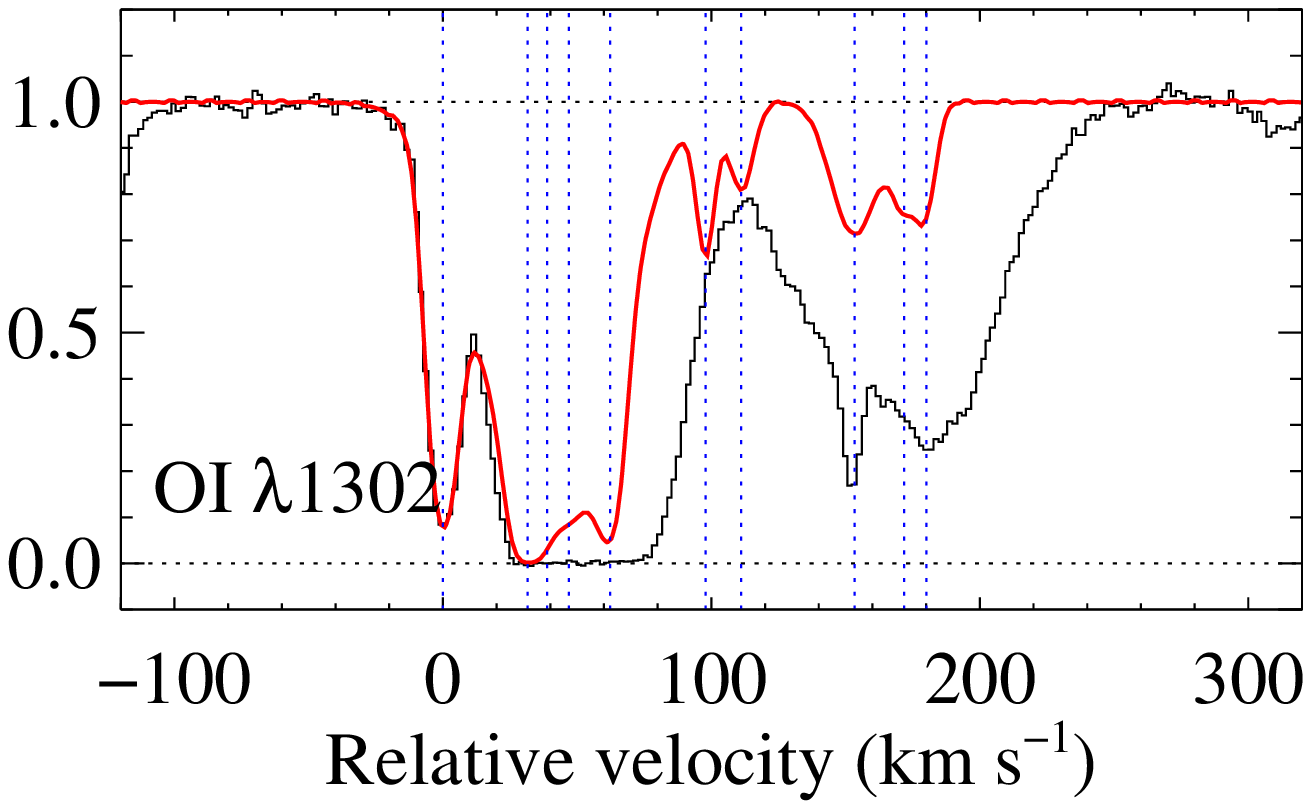}\\
\end{tabular}
\renewcommand{\tabcolsep}{6pt}
\caption{Velocity plots of \DI, \OI, and \SiII\ lines (black). The best-fit synthetic spectrum is superimposed in red. Vertical lines indicate the position of the different 
components. The profiles for \DI$\lambda$925 and \DI$\lambda$972 were computed using the derived parameters 
but were not used in the fitting process, the former being contaminated by absorption from the \lya\ 
forest and the latter being saturated. We also note that the vertical lines and the velocity scale of the 
\OI$\lambda$988 panel correspond to the strongest (\OI$\lambda$988.7) of the three transitions that produce 
the profile. \label{fit}}
\end{figure}

We focus on the $\zabs=2.621$ DLA where we detected clear, non-saturated, and unblended \DI\ absorption 
lines at $\zabs=2.62102$, which defines the zero of our adopted velocity scale (see Fig.~\ref{fit}). This 
corresponds to the bluest component of the \OI\ profile, that spans $\sim$~180~\kms. \DI\ lines 
associated with the other ($v>0$~\kms) components are in turn lost within the \HI\ profile. The total 
neutral hydrogen column density, $\log N(\HI)=20.45\pm0.10$, was measured by fitting the damped \lya\ and 
\lyb\ lines \citep[][]{Ledoux06a, Lopez03}. Since the values from the two studies (also performed using 
different instruments) agree within 0.02~dex, we did not attempt to redo this fit and adopt the above 
$N(\HI)$-value. 

\begin{figure}[t]
\centering
\includegraphics[bb = 60 202 510 399,clip=,width=0.94\hsize]{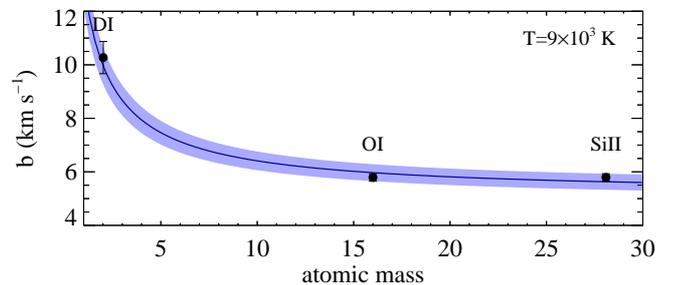}
\caption{Doppler parameter as a function of atomic mass. The solid curve and shaded area correspond to the 
best fit function $b^2= 2kT/m + b_{\rm turb}^2$ 
and 1\,$\sigma$ range with 
$b_{\rm turb}=5.2\pm0.2$~\kms and $T=8800\pm1500$~K. \label{bb}}
\end{figure}

We use standard multi-component 
Voigt-profile fitting to derive the column densities of the individual 
components, after normalising the spectrum locally around each line of interest. We simultaneously fitted 
five \OI\ transitions (at $\lambda=$ 988.5, 988.6, 988.7, 1039, and 1302) and the \SiII$\lambda$1526 
transition over the whole profile, together with \DI$\lambda$937 and \DI$\lambda$949 in the bluest component. 
The \SiII$\lambda$1526 profile is located outside the \lya\ forest, non-affected by blends, and is 
characterised by non-saturated components and a high S/N ratio. 
We used this transition as a reference to identify the individual components for the absorption 
lines located in the \lya\ forest and to get a first guess of the redshifts and Doppler parameters. During 
the fitting process, the Doppler parameters of \OI\ and \SiII\ were assumed to be equal. This is the usual 
assumption in the field and corresponds to a Doppler parameter mostly dominated by turbulent motions. 
However, because deuterium is a very light species, the contribution of thermal broadening, 
$b_{\rm th}=\sqrt{2kT/m}$ (where $k$ is the Boltzmann constant, $m$ the atomic mass and $T$ the temperature of 
the gas), is not negligible. Therefore, the Doppler parameter for atomic deuterium has been left free. 
The best fit is obtained for $b(\DI)=10.1$~\kms\ while $b=5.8$~\kms\ for \OI\ and \SiII. This corresponds 
to a kinetic temperature of the gas $T=8800 \pm 1500$~K (see Fig.~\ref{bb}), as expected in Warm Neutral 
Medium \citep{Petitjean92, Wolfire95}. Temperature measurements based on line widths have been difficult 
in DLAs \citep[see][]{Carswell12} because of their components' blending, and are thus generally limited to 
narrow lines which sample cold material much less representative of the typical DLAs \citep{Petitjean00}.

The result of the Voigt-profile fitting is shown on Fig.~\ref{fit} and the corresponding column densities 
in Table~\ref{colden}. Formal errors given by fitting procedures ($\sigma_{\rm fit}$) may underestimate 
the parameters uncertainties. 
Therefore, we also performed several fits after shifting the 
continuum by $\pm 0.5 \sigma$ (where $\sigma$ is the local flux error around each line) and take the 
difference in extreme values as the error due to uncertainties in the continuum placement, 
$\sigma_{\rm cont}$. Errors quoted in the table are estimated as 
$\sigma_{\rm tot}^2=\sigma_{\rm fit}^2+\sigma_{\rm cont}^2$.

\begin{table*}
\centering
\caption{Parameters of the modelled absorption profile \label{colden}}
\begin{tabular}{c c c c c c}
\hline
\hline
 $z$     &   $v$      &   $b$                                             &  \multicolumn{3}{c}{{\large \strut}$\log N$~(\cmsq) } \\
\cline{4-6}
         & (\kms)     & ~(\kms)                                           &    \OI\          &   \DI\         &   \HI \tablefootmark{a}  \\  
\hline                                                                                                                                       
{\large \strut}2.621022 & 0 & 5.8$\pm$0.1 (\OI) / 10.3$\pm$0.6 (\DI)      &   14.22$\pm$0.02 & 14.96$\pm$0.03 &    19.52                \\   
2.621403 & $+$31              & 6.0$\pm$0.3                               &   14.59$\pm$0.04 & --             &    19.89                \\   
2.621492 & $+$39              &28.9$\pm$1.1                               &   14.71$\pm$0.03 & --             &    20.01                \\   
2.621589 & $+$47              &11.0$\pm$2.0                               &   14.05$\pm$0.16 & --             &    19.35                \\   
2.621774 & $+$62              & 5.3$\pm$0.5                               &   14.14$\pm$0.07 & --             &    19.44                \\   
2.622204 & $+$98              & 3.6$\pm$1.4                               &   13.29$\pm$0.09 & --             &    18.59                \\   
2.622364 & $+$111             & 5.3$\pm$3.4                               &   13.14$\pm$0.15 & --             &    18.44                \\   
2.622876 & $+$153             &11.1$\pm$1.2                               &   13.62$\pm$0.11 & --             &    18.92                \\   
2.623095 & $+$172             & 6.0$\pm$2.1                               &   13.23$\pm$0.20 & --             &    18.53                \\   
2.623196 & $+$180             & 4.1$\pm$0.5                               &   13.12$\pm$0.25 & --             &    18.42                \\   
\hline                                                                                                                                                         
\multicolumn{3}{c}{{\large \strut}Total}                                  &   15.15$\pm$0.02 & --             &    20.45$\pm$0.10   \\       
\hline
\end{tabular}
\tablefoot{
\tablefoottext{a}{Total $N(\HI)$ from \citet{Ledoux06a}, distributed over the different components by 
scaling to $N(\OI)$.}
}
\end{table*}

It is not possible to directly measure the \HI\ column density in each individual component but we can reasonably 
assume that $N(\HI)$ scales directly with $N(\OI)$ \citep[see][]{Timmes97}. \OI, \HI\, and \DI\ are locked 
by favourable charge exchange reactions \citep[e.g.][]{Jenkins00a} so we do not require any 
photo-ionisation correction. Our assumption therefore simply corresponds to that of a uniform metallicity
throughout the profile\footnote{Strictly speaking, we do not even need the metallicity to be equal in 
{\sl all} the components, but that the metallicity in the $v=0$~\kms\ component equals the mean overall 
value.} (here [O/H]~=~-1.99, with respect to the solar value from \citealt{Lodders03}). Indeed, the warm 
phase of DLAs shows a high level of chemical uniformity \citep{Prochaska03a,Rodriguez06}, though we 
caution that this has not yet been tested at metallicities below 1/50$^{\rm th}$ solar. During the fitting 
process, we therefore modelled the \HI\ profiles by scaling the column densities with $N(\OI)$. The \HI\ 
$b$-values are highly degenerate, in particular in the central components, but consistent with those of 
\OI\ with additional thermal broadenings of $\sim 10^4$~K. The modelled \HI\ Ly-$\gamma$, Ly-$\delta$, 
Ly-$\epsilon$ and Ly-8 profiles provide a very satisfactory fit to the data, further supporting our above 
assumption. 

\section{Discussion \label{discussion}}

Chemical evolution models \citep[e.g.][]{Romano06} predict very little astration of deuterium at the 
metallicity measured here, [O/H]~$\approx -2$, so we can expect the deuterium abundance towards CTQ\,247 
to be very close to the primordial value. The D and O abundances are expected to be anti-correlated, 
because the former is destroyed in stars while the latter is produced during stellar nucleosynthesis. Here, 
the (D/O) ratio is much higher than what is measured locally, but also higher than several measurements 
at high-$z$ (see \citealt{Hebrard03} and references therein). Therefore, this direct 
measurement alone 
already indicates very little astration.  Furthermore, the lack of odd-even effect 
\citep[][]{Arnett71} in this low-metallicity DLA indicates a low, quiescent star-formation 
history in the system \citep{Lopez03}. Indeed, we get (D/H)~$=2.8^{+0.8}_{-0.6} \times10^{-5}$, which agrees 
remarkably well with the primordial value, (D/H)$_{\rm p} = (2.59\pm0.15)\times10^{-5}$, predicted by 
standard BBN \citep{Coc12} using $100\Omega_b h^2=2.249$ from the study of cosmic microwave background 
(CMB) anisotropies \citep{Komatsu11}. 

\begin{figure}
\centering
\includegraphics[bb = 68 179 500 570,clip=,width=\hsize]{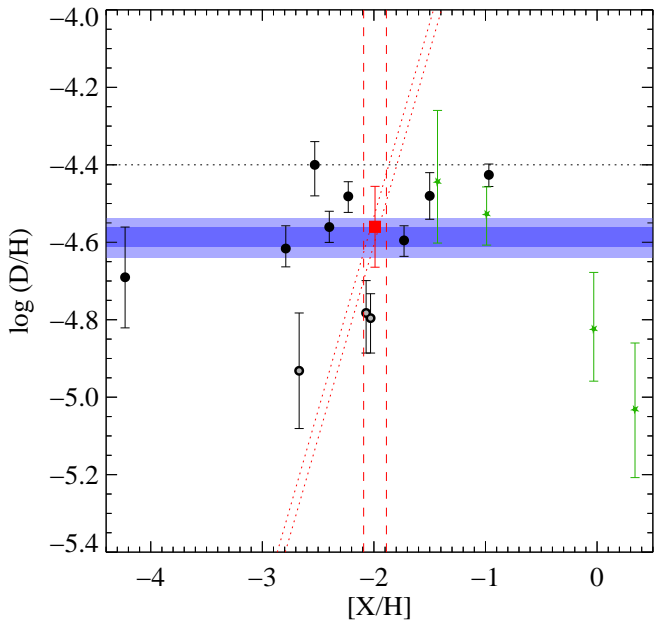}
\caption{High-redshift D/H values as a function of metallicity ([O/H] or [Si/H]). Our measurement towards 
CTQ\,247 is shown as a red square. The red dotted lines (resp. vertical dashed lines) represent our 
constraint on (D/O) (resp. [O/H]). Other measurements from the literature are shown as filled 
circles. Grey-filled circles have cautionary remarks in the text 
\citep[from left to right:][]{Srianand10,Crighton04,Pettini01}. Green stars represent HD/2H$_2$ 
measurements. The blue stripes represent the 1\,$\sigma$ and 2\,$\sigma$ prediction on (D/H)$_{\rm p}$ from 
\citet{Coc12} using standard BBN and $\Omega_b h^2$ from WMAP7 \citep[][]{Komatsu11}. The horizontal dotted 
line represent the primordial value proposed by \citet{Olive12}, see text.
\label{dhall}}
\end{figure}

In Fig.~\ref{dhall}, we compare our abundance measurement to that in other high-redshift quasar absorption 
systems \citep[][]{Burles98a, Burles98b, Crighton04, Fumagalli11, Kirkman03, Levshakov02, Omeara01, 
Omeara06, Pettini01, Pettini08, Srianand10}. The scatter between the different measurements is larger than 
the published errors. This is probably the consequence of underestimated uncertainties. 
Indeed, the robustness of each (D/H) measurement has been long discussed in the literature. For example, the 
value obtained by \citet{Crighton04} is dependent on the description of the gas kinematics and the 
error on N(\HI) is very likely underestimated \citep[][]{Omeara06}. \citet{Pettini08} also cautioned that the 
measurement towards Q\,2206$-$199 \citep{Pettini01} may suffer from a poor 
determination of $N(\DI)$ because of the low S/N ratio and resolution achieved\footnote{This system 
has been observed using HST-STIS while others were observed using echelle spectrographs on 8-10\,m class 
telescopes.}.
Overall, (D/H) ratios tend to be slightly higher than the standard BBN prediction, with a 
weighted mean $\avg{({\rm D/H})}=2.94^{+0.39}_{-0.34} \times 10^{-5}$. The same mean was obtained by \citet{Pettini08}, 
after removing values they considered not robust-enough and rescaling the errors to 
make them consistent with the observed dispersion.  These authors subsequently used the (D/H) ratios to slightly 
correct downwards $\Omega_b$ from WMAP. Yet, the study of CMB anisotropies now provides very robust 
estimation of cosmological parameters \citep{Komatsu11}, in agreement with baryon acoustic oscillation 
measurements \citep[][]{Percival10}. It seems therefore more appropriate to seek an alternative 
explanation for the high (D/H) values, if these were further confirmed. 

\citet{Olive12} recently proposed cosmological evolution models that alter the nuclear processes during or 
right after BBN and which are able to solve the $^7$Li problem while leading to a higher (D/H)$_{\rm p}$, 
possibly more consistent with the current data (horizontal dotted line on Fig.~\ref{dhall}). It is then 
still possible that the astration of deuterium decreases the (D/H) ratio locally without necessarily being 
accompanied by an increased oxygen production (as suggested by \citealt{Fields01} and possibly observed by 
\citealt{Srianand10}\footnote{Note however that the $N(\HI)$ measurement is complex in this system owing to 
strong blending of the damped lines with other systems.}). Indeed, the astration efficiency depends on the 
relative stellar populations and the ratio of gas mass to the total galactic mass 
\citep[e.g.][]{Vangioni-Flam88,Vangioni11}. Local astration could therefore explain the non-primordial 
(D/H) ratios, even at low metallicity.

Before further discussing the chemical evolution of deuterium abundance, we note that metallicities are 
difficult to establish precisely, in particular if elements other than \OI\ are used for low $N(\HI)$ 
systems since ionisation corrections can be uncertain. For example, the extremely low-metallicity claimed 
by \citet{Fumagalli11} might be underestimated because of the delicate assumption on the ionisation 
parameter. Some stellar processing of the gas is thus not completely excluded.
No decrease of the (D/H) ratio with metallicity is observed at high redshift for the full range probed 
by direct (D/H) measurements up to one tenth solar (see Fig.~\ref{dhall}), in agreement with standard 
chemical evolution models \citep[e.g.][]{Romano06}. The high-metallicity end 
could however be probed at high-$z$ in molecular-rich DLAs using [S,Zn/H] and the HD/2H$_2$ ratio 
\citep{Noterdaeme08hd,Noterdaeme10co}. As expected, the corresponding abundances are significantly lower 
than the primordial value\footnote{HD/2H$_2$ values are however above what is expected in closed-box models 
and could then indicate significant accretion of gas from the intergalactic medium, see discussion in 
\citet{Noterdaeme08hd}.}. Although the difficulties due to line identification and small numbers of 
available transitions do not apply in the case of HD-based measurements, poorly known HD-chemistry and 
self-shielding effects may alter these measurements \citep[][]{Balashev10}. Nevertheless, measurements 
performed at lower metallicities (around -1.4, \citealt{Ivanchik10} and -1, \citealt{Balashev10}) are very 
consistent with atomic-based measurements. 
The HD technique could then be a complementary way to measure the primordial value at [X/H]~$\le-1$, while 
giving constraints on galactic chemical evolution at solar metallicities. 

\section{Conclusion}
We report a new measurement of deuterium abundance in a low-metallicity DLA. The observed (D/H) value 
agrees remarkably well with the primordial value predicted by standard Big-Bang Nucleosyhthesis using the 
baryon-to-photon ratio derived independently from CMB anisotropies. As noticed by several authors 
\citep[e.g.][]{Pettini08,Ivanchik10,Srianand10,Olive12}, the dispersion of other available measurements is 
larger than the published errors. This could be either observational or physical, where high values 
could possibly be explained by cosmological evolution models including post-BBN processing of the light 
elements, and where low values would be explained by local astration. Accurately measuring the abundance of 
deuterium at high redshift is clearly a difficult task and each measurement suffers from its own 
uncertainties. It is therefore important to obtain new high-redshift measurements --of both (D/H) and 
(D/O)-- in systems with different chemical enrichment, for a better understanding of deuterium production 
during BBN and its astration due to stellar processing in the course of galactic chemical evolution.

\begin{acknowledgements}
We thank the referee for helpful comments and suggestions and Elisabeth Vangioni for useful 
discussions on the BBN. SL was supported by FONDECYT grant number 1100214 and VD by CONICYT/Gemini 
Astronomy grant for Master Students at Universidad de Chile number 32100014.
\end{acknowledgements}

\bibliographystyle{aa}

\begin{thebibliography}{45}
\expandafter\ifx\csname natexlab\endcsname\relax\def\natexlab#1{#1}\fi

\bibitem[{{Adams}(1976)}]{Adams76}
{Adams}, T.~F. 1976, \aap, 50, 461

\bibitem[{{Arnett}(1971)}]{Arnett71}
{Arnett}, W.~D. 1971, \apj, 166, 153

\bibitem[{{Balashev} {et~al.}(2010){Balashev}, {Ivanchik}, \&
  {Varshalovich}}]{Balashev10}
{Balashev}, S.~A., {Ivanchik}, A.~V., \& {Varshalovich}, D.~A. 2010, Astronomy
  Letters, 36, 761

\bibitem[{{Burles} \& {Tytler}(1998{\natexlab{a}})}]{Burles98a}
{Burles}, S. \& {Tytler}, D. 1998{\natexlab{a}}, \apj, 499, 699

\bibitem[{{Burles} \& {Tytler}(1998{\natexlab{b}})}]{Burles98b}
{Burles}, S. \& {Tytler}, D. 1998{\natexlab{b}}, \apj, 507, 732

\bibitem[{{Carswell} {et~al.}(2012){Carswell}, {Becker}, {Jorgenson}, {Murphy},
  \& {Wolfe}}]{Carswell12}
{Carswell}, R.~F., {Becker}, G.~D., {Jorgenson}, R.~A., {Murphy}, M.~T., \&
  {Wolfe}, A.~M. 2012, \mnras, 2703

\bibitem[{{Coc} {et~al.}(2012){Coc}, {Goriely}, {Xu}, {Saimpert}, \&
  {Vangioni}}]{Coc12}
{Coc}, A., {Goriely}, S., {Xu}, Y., {Saimpert}, M., \& {Vangioni}, E. 2012,
  \apj, 744, 158

\bibitem[{{Crighton} {et~al.}(2004){Crighton}, {Webb}, {Ortiz-Gil}, \&
  {Fern{\'a}ndez-Soto}}]{Crighton04}
{Crighton}, N.~H.~M., {Webb}, J.~K., {Ortiz-Gil}, A., \& {Fern{\'a}ndez-Soto},
  A. 2004, \mnras, 355, 1042

\bibitem[{{Epstein} {et~al.}(1976){Epstein}, {Lattimer}, \&
  {Schramm}}]{Epstein76}
{Epstein}, R.~I., {Lattimer}, J.~M., \& {Schramm}, D.~N. 1976, \nat, 263, 198

\bibitem[{{Fields} {et~al.}(2001){Fields}, {Olive}, {Silk}, {Cass{\'e}}, \&
  {Vangioni-Flam}}]{Fields01}
{Fields}, B.~D., {Olive}, K.~A., {Silk}, J., {Cass{\'e}}, M., \&
  {Vangioni-Flam}, E. 2001, \apj, 563, 653

\bibitem[{{Fumagalli} {et~al.}(2011){Fumagalli}, {O'Meara}, \&
  {Prochaska}}]{Fumagalli11}
{Fumagalli}, M., {O'Meara}, J.~M., \& {Prochaska}, J.~X. 2011, Science, 334,
  1245

\bibitem[{{H{\'e}brard} \& {Moos}(2003)}]{Hebrard03}
{H{\'e}brard}, G. \& {Moos}, H.~W. 2003, \apj, 599, 297

\bibitem[{{Horne}(1986)}]{Horne86}
{Horne}, K. 1986, \pasp, 98, 609

\bibitem[{{Ivanchik} {et~al.}(2010){Ivanchik}, {Petitjean}, {Balashev},
  {Srianand}, {Varshalovich}, {Ledoux}, \& {Noterdaeme}}]{Ivanchik10}
{Ivanchik}, A.~V., {Petitjean}, P., {Balashev}, S.~A., {et~al.} 2010, \mnras,
  404, 1583

\bibitem[{{Jenkins} {et~al.}(2000){Jenkins}, {Oegerle}, {Gry}, {Vallerga},
  {Sembach}, {Shelton}, {Ferlet}, {Vidal-Madjar}, {York}, {Linsky}, {Roth},
  {Dupree}, \& {Edelstein}}]{Jenkins00a}
{Jenkins}, E.~B., {Oegerle}, W.~R., {Gry}, C., {et~al.} 2000, \apjl, 538, L81

\bibitem[{{Kirkman} {et~al.}(2003){Kirkman}, {Tytler}, {Suzuki}, {O'Meara}, \&
  {Lubin}}]{Kirkman03}
{Kirkman}, D., {Tytler}, D., {Suzuki}, N., {O'Meara}, J.~M., \& {Lubin}, D.
  2003, \apjs, 149, 1

\bibitem[{{Komatsu} {et~al.}(2011){Komatsu}, {Smith}, {Dunkley}, {Bennett},
  {Gold}, {Hinshaw}, {Jarosik}, {Larson}, {Nolta}, {Page}, {Spergel},
  {Halpern}, {Hill}, {Kogut}, {Limon}, {Meyer}, {Odegard}, {Tucker}, {Weiland},
  {Wollack}, \& {Wright}}]{Komatsu11}
{Komatsu}, E., {Smith}, K.~M., {Dunkley}, J., {et~al.} 2011, \apjs, 192, 18

\bibitem[{{Ledoux} {et~al.}(2006){Ledoux}, {Petitjean}, {Fynbo}, {M{\o}ller},
  \& {Srianand}}]{Ledoux06a}
{Ledoux}, C., {Petitjean}, P., {Fynbo}, J.~P.~U., {M{\o}ller}, P., \&
  {Srianand}, R. 2006, \aap, 457, 71

\bibitem[{{Ledoux} {et~al.}(2003){Ledoux}, {Petitjean}, \&
  {Srianand}}]{Ledoux03}
{Ledoux}, C., {Petitjean}, P., \& {Srianand}, R. 2003, \mnras, 346, 209

\bibitem[{{Levshakov} {et~al.}(2002){Levshakov}, {Dessauges-Zavadsky},
  {D'Odorico}, \& {Molaro}}]{Levshakov02}
{Levshakov}, S.~A., {Dessauges-Zavadsky}, M., {D'Odorico}, S., \& {Molaro}, P.
  2002, \apj, 565, 696

\bibitem[{{Lodders}(2003)}]{Lodders03}
{Lodders}, K. 2003, \apj, 591, 1220

\bibitem[{{L\'opez} \& {Ellison}(2003)}]{Lopez03}
{L\'opez}, S. \& {Ellison}, S.~L. 2003, \aap, 403, 573

\bibitem[{{L\'opez} {et~al.}(2001){L\'opez}, {Maza}, {Masegosa}, \&
  {Marquez}}]{Lopez01}
{L\'opez}, S., {Maza}, J., {Masegosa}, J., \& {Marquez}, I. 2001, \aap, 366,
  387

\bibitem[{{Noterdaeme} {et~al.}(2010){Noterdaeme}, {Petitjean}, {Ledoux},
  {L{\'o}pez}, {Srianand}, \& {Vergani}}]{Noterdaeme10co}
{Noterdaeme}, P., {Petitjean}, P., {Ledoux}, C., {et~al.} 2010, \aap, 523, A80

\bibitem[{{Noterdaeme} {et~al.}(2008){Noterdaeme}, {Petitjean}, {Ledoux},
  {Srianand}, \& {Ivanchik}}]{Noterdaeme08hd}
{Noterdaeme}, P., {Petitjean}, P., {Ledoux}, C., {Srianand}, R., \& {Ivanchik},
  A. 2008, \aap, 491, 397

\bibitem[{{Olive} {et~al.}(2012){Olive}, {Petitjean}, {Vangioni}, \&
  {Silk}}]{Olive12}
{Olive}, K.~A., {Petitjean}, P., {Vangioni}, E., \& {Silk}, J. 2012, \mnras,
  submitted [arXiv:1203.5701]

\bibitem[{{O'Meara} {et~al.}(2006){O'Meara}, {Burles}, {Prochaska}, {Prochter},
  {Bernstein}, \& {Burgess}}]{Omeara06}
{O'Meara}, J.~M., {Burles}, S., {Prochaska}, J.~X., {et~al.} 2006, \apjl, 649,
  L61

\bibitem[{{O'Meara} {et~al.}(2001){O'Meara}, {Tytler}, {Kirkman}, {Suzuki},
  {Prochaska}, {Lubin}, \& {Wolfe}}]{Omeara01}
{O'Meara}, J.~M., {Tytler}, D., {Kirkman}, D., {et~al.} 2001, \apj, 552, 718

\bibitem[{{Percival} {et~al.}(2010){Percival}, {Reid}, {Eisenstein}, {Bahcall},
  {Budavari}, {Frieman}, {Fukugita}, {Gunn}, {Ivezi{\'c}}, {Knapp}, {Kron},
  {Loveday}, {Lupton}, {McKay}, {Meiksin}, {Nichol}, {Pope}, {Schlegel},
  {Schneider}, {Spergel}, {Stoughton}, {Strauss}, {Szalay}, {Tegmark},
  {Vogeley}, {Weinberg}, {York}, \& {Zehavi}}]{Percival10}
{Percival}, W.~J., {Reid}, B.~A., {Eisenstein}, D.~J., {et~al.} 2010, \mnras,
  401, 2148

\bibitem[{{Petitjean} {et~al.}(1992){Petitjean}, {Bergeron}, \&
  {Puget}}]{Petitjean92}
{Petitjean}, P., {Bergeron}, J., \& {Puget}, J.~L. 1992, \aap, 265, 375

\bibitem[{{Petitjean} {et~al.}(2000){Petitjean}, {Srianand}, \&
  {Ledoux}}]{Petitjean00}
{Petitjean}, P., {Srianand}, R., \& {Ledoux}, C. 2000, \aap, 364, L26

\bibitem[{{Pettini} \& {Bowen}(2001)}]{Pettini01}
{Pettini}, M. \& {Bowen}, D.~V. 2001, \apj, 560, 41

\bibitem[{{Pettini} {et~al.}(2008){Pettini}, {Zych}, {Murphy}, {Lewis}, \&
  {Steidel}}]{Pettini08}
{Pettini}, M., {Zych}, B.~J., {Murphy}, M.~T., {Lewis}, A., \& {Steidel}, C.~C.
  2008, \mnras, 391, 1499

\bibitem[{{Prochaska}(2003)}]{Prochaska03a}
{Prochaska}, J.~X. 2003, \apj, 582, 49

\bibitem[{{Prodanovi{\'c}} \& {Fields}(2003)}]{Prodanovic03}
{Prodanovi{\'c}}, T. \& {Fields}, B.~D. 2003, \apj, 597, 48

\bibitem[{{Rodr{\'{\i}}guez} {et~al.}(2006){Rodr{\'{\i}}guez}, {Petitjean},
  {Aracil}, {Ledoux}, \& {Srianand}}]{Rodriguez06}
{Rodr{\'{\i}}guez}, E., {Petitjean}, P., {Aracil}, B., {Ledoux}, C., \&
  {Srianand}, R. 2006, \aap, 446, 791

\bibitem[{{Romano} {et~al.}(2006){Romano}, {Tosi}, {Chiappini}, \&
  {Matteucci}}]{Romano06}
{Romano}, D., {Tosi}, M., {Chiappini}, C., \& {Matteucci}, F. 2006, \mnras,
  369, 295

\bibitem[{{Srianand} {et~al.}(2010){Srianand}, {Gupta}, {Petitjean},
  {Noterdaeme}, \& {Ledoux}}]{Srianand10}
{Srianand}, R., {Gupta}, N., {Petitjean}, P., {Noterdaeme}, P., \& {Ledoux}, C.
  2010, \mnras, 405, 1888

\bibitem[{{Steigman}(1994)}]{Steigman94}
{Steigman}, G. 1994, \mnras, 269, L53

\bibitem[{{Steigman}(2007)}]{Steigman07a}
{Steigman}, G. 2007, Annual Review of Nuclear and Particle Science, 57, 463

\bibitem[{{Timmes} {et~al.}(1997){Timmes}, {Truran}, {Lauroesch}, \&
  {York}}]{Timmes97}
{Timmes}, F.~X., {Truran}, J.~W., {Lauroesch}, J.~T., \& {York}, D.~G. 1997,
  \apj, 476, 464

\bibitem[{{Vangioni} {et~al.}(2011){Vangioni}, {Silk}, {Olive}, \&
  {Fields}}]{Vangioni11}
{Vangioni}, E., {Silk}, J., {Olive}, K.~A., \& {Fields}, B.~D. 2011, \mnras,
  413, 2987

\bibitem[{{Vangioni-Flam} \& {Audouze}(1988)}]{Vangioni-Flam88}
{Vangioni-Flam}, E. \& {Audouze}, J. 1988, \aap, 193, 81

\bibitem[{{Wolfire} {et~al.}(1995){Wolfire}, {Hollenbach}, {McKee}, {Tielens},
  \& {Bakes}}]{Wolfire95}
{Wolfire}, M.~G., {Hollenbach}, D., {McKee}, C.~F., {Tielens}, A.~G.~G.~M., \&
  {Bakes}, E.~L.~O. 1995, \apj, 443, 152

\bibitem[{{York}(2002)}]{York02}
{York}, D.~G. 2002, \planss, 50, 1251

\end{thebibliography}

\end{document}